\documentclass[aps,prd,nofootinbib,preprint,superscriptaddress]{revtex4}  
\usepackage{epsfig} 
\begin{document}

\title{The Isgur-Karl model revisited}

\author{Leonardo Galeta} 
\affiliation{Departamento de F\'{\i}sica,
FCEyN, Universidad de Buenos Aires, Ciudad Universitaria, Pab.1, (1428)
Buenos Aires, Argentina} 
\author{Dan Pirjol} 
\affiliation{National
Institute for Physics and Nuclear Engineering, Department of Particle
Physics, 077125 Bucharest, Romania}
\author{Carlos Schat} 
\affiliation{Department of Physics and Astronomy, Ohio University, Athens, Ohio 45701, USA}
\affiliation{Departamento de F\'{\i}sica, FCEyN,
Universidad de Buenos Aires, Ciudad Universitaria, Pab.1, (1428) Buenos
Aires, Argentina} 

\date{\today}

\begin{abstract} 
We show how to match the Isgur-Karl model to the spin-flavor quark 
operator expansion used in the $1/N_c$ studies of the non-strange 
negative parity $L=1$ excited baryons. Using the transformation properties 
of states and interactions under the permutation group $S_3$ we are 
able to express the operator coefficients as overlap integrals, 
without making any assumption on the spatial dependence of the quark 
wave functions. The general mass operator leads to parameter free mass 
relations and constraints on the mixing angles that are valid beyond 
the usual harmonic oscillator approximation.
The Isgur-Karl model with harmonic oscillator wave functions provides a simple counterexample 
that demonstrates explicitly that  the alternative 
operator basis for the $1/N_c$ expansion for excited baryons recently 
proposed by Matagne and Stancu is incomplete. 
	 
\end{abstract}


\maketitle

\section{Introduction}

Excited baryons are the natural playground to test 
the spin-flavor structure of quark interactions in the low-energy
regime and provide useful information about the nonperturbative aspects
of quantum chromodynamics.  
A simple model used to study the masses and mixing angles of
excited baryons is the Isgur-Karl (IK) model \cite{Isgur:1977ef}.
In this model the interaction Hamiltonian of two quarks contains two
components: a contact spin-spin term and a tensor interaction. This is
an approximation to the Breit interaction of two quarks mediated by
one-gluon exchange \cite{De Rujula:1975ge} (the OGE model), obtained by
neglecting the spin-orbit interaction. The physical motivation for
neglecting the spin-orbit interaction is debatable; we will assume it
from the start as defining the model considered here.

The predictions of the IK model have been
obtained assuming a  harmonic oscillator basis for the orbital wave functions
 \cite{Isgur:1977ef}. 
With this assumption the
model is very predictive: the entire mass spectrum of the $L=1$ negative
parity baryons is determined in terms of two free parameters, and the mixing
angles are independent of the hadron masses. 

In this paper we concentrate on these states and show how to rewrite the IK 
model predictions in an equivalent 
way, constructing its effective mass operator in terms of a spin-flavor quark 
operator expansion. This type of operator expansion is used in a systematic 
manner in the $1/N_c$ studies of excited baryons \cite{Goity:1996hk},\cite{Carlson:1998vx}, 
where more general spin-flavor quark-quark interactions are allowed for. 

The motivation for performing the matching of the IK model to the more 
general $1/N_c$ expansion is twofold: In the IK model the computation of  the coefficients
of the operator expansion is straightforward
and 
illustrates the connection of a model calculation with the $1/N_c$ expansion 
explicitly. The second reason is that it provides a simple counterexample that 
shows the incompleteness of  the 
alternative operator basis advocated recently by Matagne and Stancu in 
Ref.~\cite{Matagne:2006dj}. The usual basis with excited quark and core operators 
can  reproduce the 
IK predictions, while a basis of symmetric operators as proposed in 
Ref.~\cite{Matagne:2006dj} can not do it.

To compute the matching we use the method proposed in a recent paper
\cite{Pirjol:2007ed}, which considers the transformation properties of
the states and operators under $S_3$, the permutation group of three objects
acting on the spatial and spin-flavor degrees of freedom. 
Using these transformation properties under $S_3$ the 
coefficients of the operator expansion can be expressed
as overlap integrals, without making 
any assumption on the spatial dependence of the quark wave functions. This 
allows one to obtain mass relations and constraints on the mixing angles 
that are valid beyond the harmonic oscillator approximation of the IK model.

Examining the transformation properties 
of states and operators under the permutation group $S_3 $ also allows to count the 
number of unknown parameters (reduced matrix elements) that follow from a specific form 
of the quark-quark interaction, as was already discussed in Ref.~\cite{Collins:1998ny}. 
In the IK model the spatial and spin-flavor components of the  spin-spin and tensor interactions 
are both two-body symmetric interactions of dimension $\mathbf{3}$ that decompose 
as\footnote{In the following S, MS and A are the 
symmetric, mixed symmetric and antisymmetric irreps of $S_3$ of 
dimensions one, two  and one respectively.}  ${\rm S \oplus MS}$ under
$S_3$. The spatial and spin-flavor part of the $L=1$ excited baryons states we consider 
here transform both as MS. In the matrix elements only operators 
that transform as irreps contained in the decomposition of $ {\rm MS \otimes MS}$  can contribute.
S and MS appear once in the decomposition of 
${\rm MS \otimes MS=S \oplus MS \oplus A}$, which indicates that there will be two unknown 
reduced matrix elements 
for each of the spin-spin and tensor interactions. The unit operator coming 
from the confinement potential is also present and transforms as S under $S_3$. This leads 
to five unknowns in the most general case. We show later that for a spin-spin 
contact interaction the two reduced matrix elements are related and the 
most general mass operator depends on four unknown coefficients. 
 In the particular case of the harmonic oscillator approximation taken in the 
original formulation of the IK model, all the reduced matrix elements that 
contribute to the splittings 
are related and can be parameterized by a single parameter.

The paper is organized as follows. In Sec.~\ref{states} we present the 
excited baryon states, in Sec.~\ref{IKV} we discuss the general form of the 
matrix elements using $S_3$ and  in Sec.~\ref{IKVpred} we give the 
general mass relations and constraints on the mixing angles.
In Sec.~\ref{IKho} we discuss the predictions 
of the IK model with harmonic oscillator wave functions. Finally, 
in Sec.~\ref{ncexp} we discuss on hand of the IK model that the 
inclusion of excited and core quark operators is needed and in Sec.~\ref{concl}
we give our conclusions.

\section{The states}
\label{states}

The $L=1$ quark model states for the excited baryons we will consider here, have both the spatial and the spin-flavor
wave functions transforming in the mixed symmetric irreducible
representation of $S_3$. A two-dimensional basis for the representation
can be chosen as $\chi_i(\vec r_1,\vec r_2,\vec r_3)$, for the spatial wave functions, and
$\phi_j$ for the spin-flavor wave functions, with $i,j=2,3$. The total
wave function $|B\rangle$ is the tensor product of the
spatial-spin-flavor wave functions which is completely symmetric (and antisymmetric in 
color).

A special choice of the MS basis wave functions was adopted in
Ref.~\cite{Pirjol:2007ed} (from here on referred as I), motivated by computational ease in the
arbitrary $N_c$ case. This choice is defined by the transformation
properties of the basis under permutations, given by Eqs.~(6)-(8) in I.
For $N_c=3$ the defining properties of the basis states are
\begin{eqnarray}\label{chibasis} 
& &P_{12} \chi_2 = - \chi_2\,,
\qquad\,\,\,\, P_{12} \chi_3 = \chi_3 - \chi_2 \nonumber \,, \\ 
& &P_{13} \chi_2 =
\chi_2 - \chi_3\,, \quad P_{13} \chi_3 = -\chi_3 \,, \\ 
& &P_{23} \chi_2 =
\chi_3\,, \qquad\quad\, P_{23} \chi_3 = \chi_2 \,. \nonumber
\end{eqnarray} 
We will relate this basis to the $\rho,\lambda$ basis commonly used in the IK model in Section \ref{IKho}.
The  basis of spin-flavor wave functions $\phi_j$ can be chosen to have the same properties under 
permutations as $\chi_i$. An explicit example for the $\phi_j$
basis can be found in Appendix B of reference I for the $N_{5/2}(1675)$ state.  We
will use the same basis here, which will allow us to use the results for
matrix elements derived in I.

With the basis choice defined by Eq.~(\ref{chibasis}), the complete
baryon wave function is given by Eq.~(10) of I  
\begin{eqnarray}
|B(J,m_J)\rangle = \frac{\sqrt{2}}{3}\sum_{i,j=2}^3 \chi_i(L,m_L) \phi_j(S,m_S,I,I_3) 
\left( \begin{array}{cc} 1 &
-\frac12 \\ -\frac12 & 1 \\ 
\end{array} \right)_{ij} \langle J, m_J|L,S; m_L,m_S\rangle \,.  
\end{eqnarray}
We made here explicit the spin quantum numbers of the spatial $\chi_i$ and
spin-flavor $\phi_j$ states, although for reasons of simplicity they will
be omitted in the following. We also included a normalization factor that 
normalizes the states as $\langle B |B\rangle = 1.$ These spatial 
(and similarly the spin-flavor) MS basis is normalized as
$\langle \chi_i|\chi_j\rangle = 2$, if $i=j$, and 
$\langle \chi_i|\chi_j\rangle = 1$ if $i\neq j$.
It is easy to verify using Eqs.~(\ref{chibasis}) that the state
$|B\rangle$ is indeed invariant under any permutation of two quarks.

The quark spin can be $S=1/2, 3/2$, which is combined with the orbital
angular momentum $L=1$ to give the following $N$ states: two states with
$J=1/2$ denoted $N_{1/2}, N'_{1/2}$, two states $J=3/2$ denoted
$N_{3/2}, N'_{3/2}$, and one state with $J=5/2$ denoted $N_{5/2}$. In
addition, there are also two $\Delta$ states, denoted as $\Delta_J$ with
$J=1/2,3/2$.

States with the same quantum numbers mix, and we define the relevant
mixing angles in the nonstrange sector as 
\begin{eqnarray} N(1535) &=&
\cos\theta_{N1} N_{1/2} + \sin\theta_{N1} N'_{1/2} \,, \\ 
N(1650) &=& -\sin\theta_{N1} N_{1/2} + \cos\theta_{N1} N'_{1/2} \,, 
\end{eqnarray} 
for
the spin-1/2 nucleons, and 
\begin{eqnarray} N(1520) &=& \cos\theta_{N3}
N_{3/2} + \sin\theta_{N3} N'_{3/2} \,, \\
N(1700) &=& -\sin\theta_{N3}
N_{3/2} + \cos\theta_{N3} N'_{3/2} \,, 
\end{eqnarray} 
for the spin-3/2
nucleons.  The quark model basis states $(N_{J},N'_{J})$ 
have quark spin $S=(1/2,3/2)$, respectively.
It is possible to bring the mixing angles into
the range $(0^\circ , 180^\circ)$ by appropriate phase redefinitions of
the physical states. We will use in the numerical analysis the hadronic
masses in Table~\ref{table1}, taken from Ref.~\cite{Amsler:2008zzb}.

\section{The mass operator of the Isgur-Karl model}
\label{IKV}

The Isgur-Karl model is defined by the quark Hamiltonian
\begin{eqnarray}
{\cal H}_{IK} = H_0 + {\cal H}_{\rm hyp}  \,, 
\end{eqnarray}
where
$H_0$ contains the confining potential and kinetic terms of the quark
fields, and is symmetric under spin and isospin. The hyperfine
interaction ${\cal H}_{\rm hyp}$ is given by 
\begin{eqnarray}\label{HIK} 
{\cal H}_{\rm hyp} = A \sum_{i<j}\Big[ \frac{8\pi}{3} \vec s_i \cdot \vec s_j
\delta^{(3)}(\vec r_{ij}) + \frac{1}{r_{ij}^3} (3\vec s_i \cdot \hat r_{ij} \
\vec s_j \cdot \hat r_{ij} - \vec s_i\cdot \vec s_j) \Big] \,, 
\end{eqnarray} 
where $A$ determines the strength of the interaction, and
$\vec r_{ij} = \vec r_i - \vec r_j$ is the distance between quarks
$i,j$.  The first term is a local spin-spin interaction, and the second
describes a tensor interaction between two dipoles. This interaction
Hamiltonian is an approximation to the gluon-exchange interaction,
neglecting the spin-orbit terms\footnote{In Ref.\cite{Isgur:1977ef} A is taken as $A=\frac{2 \alpha_S}{3 m^2}$.}. 

 In the original formulation of the 
IK model \cite{{Isgur:1977ef}} the confining forces are harmonic and 
we will refer to this model as IK-h.o. (harmonic oscillator). We will derive in the following the form 
of the mass operator without making any assumption on the form of the confining quark forces. 
We refer to this version of the model as IK-V(r). 

We obtain in the following the explicit form of the mass operator of
this model in the system of the $L=1$ negative parity baryons, following
the method based on the permutation group $S_3$ presented in I.  The
interaction Hamiltonian Eq.~(\ref{HIK}) has the general form
\begin{eqnarray}\label{Hhyp}
{\cal H}_{\rm hyp} = \sum_{i<j} {\cal R}_{ij} \cdot {\cal O}_{ij} \,, 
\end{eqnarray} 
where ${\cal R}_{ij}$ are orbital operators
acting on the coordinates of the quarks $i,j$, and ${\cal O}_{ij}$ are
spin-flavor operators. Both can also carry spatial indices, which are contracted to form a scalar in ${\cal H}_{\rm hyp}$, as indicated by the 
dot product in Eq.~(\ref{Hhyp}). 

The orbital and spin-flavor
operators for the contact and tensor interactions are
\begin{equation} \label{defRQ}
\begin{array}{ll}
R_{ij} = \frac{8\pi}{3}A\delta^{(3)}(\vec r_{ij})\,, & { O}_{ij}=   s_i \cdot s_j  \,, \\
Q_{ij}^{ab} = \frac{A}{r_{ij}^3} (3\hat r_{ij}^a \hat r_{ij}^b - \delta^{ab}) \,,\qquad &
{ O}_{ij}^{ab}= \frac12  (s_i^a s_j^b + s_i^b s_j^a)  \,,
\end{array}
\end{equation}
where $a,b$ are spatial indices. All these operators are
symmetric under the permutation of the two quark indices $i,j$, but 
belong to the reducible representation $\mathbf{3}$ under the
permutation of the three quarks.

It has been shown in I that the hadronic matrix
elements of the Hamiltonian ${\cal H}_{\rm hyp}$
can be expressed in terms of matrix elements of 
spin-flavor operators $O_i$ that are related to the decomposition of
${\cal O}_{ij}$ into irreducible representations of $S_3$, the permutation
group of three objects 
\begin{eqnarray}\label{Vqqeff} 
\langle B | {\cal H}_{\rm hyp} | B\rangle = \sum_{i} c_i \langle \Phi(SI) | O_i | \Phi(SI) \rangle \,, 
\end{eqnarray} where
the coefficients $c_i$ contain the reduced matrix elements of the
orbital operators ${\cal R}_{ij}$, and can be written in terms of
overlap integrals of the quark model wave functions.
The matrix elements of the spin-flavor operators in Eq.~(\ref{Vqqeff}) are 
a convenient way to obtain the reduced
matrix elements of the projections of ${\cal O}_{ij}$ onto irreducible
representations of $S_3^{\rm sp-fl}$. They have been computed in
I, and are taken between 
the states $|\Phi(SI)\rangle$ constructed  in
Ref.~\cite{Carlson:1998vx} as the tensor product of the ``excited''
quark 1 with a core of unexcited quarks 2,3, and projected onto the MS irrep of
spin-flavor SU(4). The advantage of this representation is that the
relevant matrix elements can be immediately read off from the tables in
Ref.~\cite{Carlson:1998vx}.

The general form of the matrix element of ${\cal H}_{\rm hyp}$ can be
taken from Eq.~(37) of I, which we repeat here for the convenience 
of the reader:
\begin{eqnarray}
\langle B | {\cal H}^{symm}| B\rangle =
\frac13 \langle {\cal R}^S \rangle \langle {\cal O}^S \rangle 
+ \frac13 \langle {\cal R}^{MS} \rangle \langle {\cal O}^{MS} \rangle \,.
\end{eqnarray}
The reduced matrix elements $\langle{\cal O}^S\rangle$ and $\langle{\cal O}^{MS}\rangle$ for the spin-spin and tensor interaction are written in terms of matrix elements of spin-flavor operators taken between the spin flavor states $|\Phi(SI)\rangle$. The corresponding expressions for arbitrary $N_c$ can be found in Eqs.(39),(42),(49),(55) of I. Here we present the $N_c=3$ expression
\begin{eqnarray}\label{ME} 
& & \langle B |
{\cal H}_{\rm hyp} |B\rangle = \frac13 \langle R_S\rangle 
\Big( \frac12\vec S^2 - \frac98 \Big) + 
\frac13 \langle R_{MS}\rangle \Big( - \vec
S^2 + 3 \vec s_1 \cdot \vec S_c + \frac94 \Big)\\ 
& &\qquad + \frac13
\langle Q_S\rangle \Big( \frac14 L_2^{ab} \{ S^a\,, S^b\} \Big) +
\frac13 \langle Q_{MS}\rangle \Big( \frac32 L_2^{ab} \{ s_1^a\,, S_c^b\} -
\frac12 L_2^{ab} \{S^a\,, S^b\} \Big) \,,  \nonumber 
\end{eqnarray} 
where the first
line corresponds to the contact term, and the second line to the tensor
term, with $L_2^{ab} = \frac12 \{ L^a, L^b\} - \frac13 L(L+1)\delta^{ab}$.
The reduced matrix elements of the orbital operators $\langle
R_S\rangle , \langle R_{MS}\rangle , \langle Q_S\rangle , \langle
Q_{MS}\rangle $ are given by (unknown) overlap integrals of the
corresponding operators with the wave functions of the states of
interest. The reduced matrix elements are defined explicitly below 
in Eq.~(\ref{R12}) for the orbital operator $R_{ij}$ appearing in 
the definition of the spin-spin interaction, and in Eq.~(\ref{Q12}) 
for the orbital operator $Q_{12}^{ab}$ appearing in the definition of 
the quadrupole interaction.

We examine now closer the structure of the orbital matrix elements. There
are three orbital operators $R_{ij}$, which transform as a combination
of S and MS under $S_3$. The symmetric projection is 
\begin{eqnarray}
R_S = R_{12} + R_{13} + R_{23} \,, 
\end{eqnarray} 
and the MS operators are
\begin{eqnarray} 
R_{MS}^2 &=& R_{13} - R_{23} \,, \\ 
R_{MS}^3 &=& R_{12} - R_{23} \,.
\end{eqnarray} 
Their matrix elements on a 2-dimensional basis of
MS wave functions $(\chi_2, \chi_3)$ with their reduced matrix elements 
defined by 
Eqs.~(34)-(36) in I, are given by 
\begin{eqnarray} 
\langle \chi_i |R_S
|\chi_j \rangle &=& \langle R_S \rangle  
\left( \begin{array}{cc} 2 & 1
\\ 1 & 2 \\ \end{array} \right)_{ij} \,, \\ 
\langle \chi_i |R_{MS}^2 |\chi_j
\rangle &=& \langle R_{MS} \rangle  \left( \begin{array}{cc} 0 & 1 \\ 1
& 1 \\ \end{array} \right)_{ij} \,, \\ 
\langle \chi_i |R_{MS}^3 |\chi_j
\rangle &=& \langle R_{MS} \rangle  \left( \begin{array}{cc} 1 & 1 \\ 1
& 0 \\ \end{array} \right)_{ij} \,.
\end{eqnarray} 
These equations can be
solved for the matrix elements of $R_{12}$, acting on quarks $1,2$, with
the result 

\begin{eqnarray}\label{calR12}
\langle \chi_i |{\cal R}_{12} |\chi_j \rangle &=&
\frac13 \left( \begin{array}{cc} 2(\langle {\cal R}_{S} \rangle+\langle {\cal R}_{MS}
\rangle) &  \langle {\cal R}_{S} \rangle+\langle {\cal R}_{MS} \rangle \\ \langle
{\cal R}_{S} \rangle+\langle {\cal R}_{MS} \rangle & 2\langle {\cal R}_{S} \rangle-\langle
{\cal R}_{MS} \rangle \\ \end{array} \right)_{ij} \,.
\end{eqnarray}

The spatial MS basis, as well as the operators,  also carry angular momentum indices. Applying the 
Wigner Eckart for SU(2) one can factor the dependence on the magnetic 
quantum numbers $m,m'$. In the 
case of a scalar operator like the spin-spin interaction one obtains:

\begin{eqnarray}\label{R12}
\langle \chi_i(1m') |R_{12} |\chi_j(1m) \rangle &=&
\frac13 \left( \begin{array}{cc} 2(\langle R_{S} \rangle+\langle R_{MS}
\rangle) &  \langle R_{S} \rangle+\langle R_{MS} \rangle \\ \langle
R_{S} \rangle+\langle R_{MS} \rangle & 2\langle R_{S} \rangle-\langle
R_{MS} \rangle \\ \end{array} \right)_{ij} \delta_{m m'} \,.
\end{eqnarray}

In the case of a tensor operator one obtains:

\begin{eqnarray}\label{Q12}
\langle \chi_i(1m') |Q_{12}^{ab} |\chi_j(1m) \rangle &=&
\frac13 \left( \begin{array}{cc} 2(\langle Q_{S} \rangle+\langle Q_{MS}
\rangle) &  \langle Q_{S} \rangle+\langle Q_{MS} \rangle \\ \langle
Q_{S} \rangle+\langle Q_{MS} \rangle & 2\langle Q_{S} \rangle-\langle
Q_{MS} \rangle \\ \end{array} \right)_{ij} 
\Big( \frac12 \{ L^a, L^b\} - \frac23 \delta^{ab} \Big)_{m',m}\nonumber\\
\end{eqnarray}

The basis for the MS orbital wave functions in I is chosen such that
$\chi_2$ satisfies $P_{12}\chi_2 = -\chi_2$, and is thus odd under a
permutation of the quarks $1,2$.  This implies that $\chi_2(r_i)$
vanishes for $r_{12} = 0$, giving
\begin{eqnarray} 
\langle \chi_2
|\delta^{(3)}(\vec r_{12}) |\chi_2 \rangle = 2(\langle R_S\rangle  + \langle
R_{MS}\rangle ) = 0\,,
\end{eqnarray} 
which  implies a relation among the
$R_S$ and $R_{MS}$  reduced matrix elements, generally valid for any
local interaction,  $\langle R_{MS} \rangle =-\langle R_{S}\rangle$. 

Using this relation in Eq.~(\ref{ME}), one finds that the most general
mass operator in the IK model depends only on three unknown orbital overlap
integrals, plus an additive constant $c_0$ related to the matrix element
of $H_0$, and can be written as 
\begin{eqnarray}\label{IKMass} 
\hat M =
c_0 + a  S_c^2 + b L_2^{ab} \{ S_c^a\,, S_c^b\}  + c L_2^{ab} \{
s_1^a\,, S_c^b\} \,,
\end{eqnarray} 
where the spin-flavor operators are
understood to act on the state $|\Phi(SI)\rangle$ constructed as a
tensor product of the core of quarks 2,3 and the `excited' quark 1.  The
coefficients are given by
\begin{eqnarray}\label{coefa}
a &=& \frac12 \langle R_S\rangle \,, \\
b &=& \frac{1}{12} \langle Q_S\rangle - \frac16 \langle Q_{MS}\rangle \,, \\
c &=& \frac16 \langle Q_S\rangle + \frac16 \langle Q_{MS}\rangle \,. \label{coefc}
\end{eqnarray}

Evaluating the matrix elements using the tables in
Ref.~\cite{Carlson:1998vx} we find the following explicit result for
the mass matrix 
\begin{eqnarray}
M_{1/2} &=& 
\left( 
\begin{array}{cc}
c_0 + a                   &  -\frac53 b + \frac{5}{6}c \\ 
-\frac53 b + \frac{5}{6}c & c_0 + 2a + \frac53(b+c)\\ 
\end{array} 
\right) \,, \\ 
M_{3/2} &=& 
\left(
\begin{array}
{cc} c_0 + a                                  &  \frac{\sqrt{10}}{6} b -\frac{\sqrt{10}}{12}c \\ 
\frac{\sqrt{10}}{6} b - \frac{\sqrt{10}}{12}c & c_0 + 2a - \frac43(b+c)\\ 
\end{array}
\right) \,, \\ 
M_{5/2}      &=& c_0 + 2a +\frac13 (b+c) \,, \\ 
\Delta_{1/2} &=& \Delta_{3/2} = c_0 + 2a \,.
\end{eqnarray} 
In the next Section we study the implications of these
results.

\section{Predictions from the IK-V(r) model}
\label{IKVpred}

The IK model makes several predictions which are independent of
the values of the overlap integrals $c_0, a,b,c$ and are valid beyond the harmonic oscillator 
approximation. 

First, the masses of the $\Delta_{1/2}$ and $\Delta_{3/2}$ states are
predicted to be equal. Experimentally, they are split by $\Delta_{3/2} -
\Delta_{1/2} = 80 \pm 50$ MeV. This mass splitting is introduced by the
spin-orbit coupling, which is neglected in the Isgur-Karl model.

Second, the splittings  $\langle \Delta\rangle - N_{5/2}$ and $\langle
N_{3/2}\rangle - \langle N_{1/2}\rangle $ are predicted to be related as
\begin{eqnarray}\label{rel1} 
\langle \Delta\rangle - N_{5/2} = \frac29
(\langle N_{3/2}\rangle - \langle N_{1/2}\rangle)\,.
\end{eqnarray}
The angular brackets denote spin-weighted averaging over the
corresponding doublets 
\begin{eqnarray} \langle \Delta\rangle &=&
\frac13 \Delta_{1/2} + \frac23 \Delta_{3/2} = 1683.3 \pm 28.5 \mbox{ MeV} \,, \\ 
\langle N_{1/2}\rangle &=& \frac12 (N(1535) + N(1650)) = 1596.5 \pm 8.2 \mbox{ MeV} \,, \\
\langle N_{3/2}\rangle &=& \frac12 (N(1520) + N(1700)) = 1610.0 \pm 25.1 
\mbox{ MeV}\,.  
\end{eqnarray} 
The experimental values of the two sides of
Eq.~(\ref{rel1}) are (in MeV) 
\begin{eqnarray} 8.3 \pm 28.9 = 3.0 \pm 5.9 \,, 
\end{eqnarray} 
which is well satisfied within errors.

\begin{table}
\begin{tabular}{|c|ccccccc|c|} 
\hline  
label & $N_{1/2}(1535)$ & $N_{1/2}(1650)$ & $N_{3/2}(1520)$ & $N_{3/2}(1700)$ & $N_{5/2}(1675)$&$\Delta_{1/2}(1620)$ & $\Delta_{3/2}(1700)$ & \ \ \ $\chi^2$ \ \ \ \\ 
\hline
PDG(2008) & $1535\pm 10$ & $1658\pm 13$ & $1520\pm 5$ & $1700\pm 50$ & $1675 \pm 5$ & $1630\pm 30$ & $1710\pm 40$ & - \\ 
\hline 
IK-V(r)  & 
$1523$ & $1659$ & $1523$ & $1693$ & $1674$ & $1678$ & $1678$ & 5.0 \\ 
\hline
IK-h.o.  &  
$1490$ & $1657$ & $1533$ & $1749$ & $1671$ & $1686$ & $1686$ & 33. \\ 
\hline
\end{tabular} 
\caption{The experimental values are taken from Ref.~\cite{Amsler:2008zzb}. IK-V(r) is the best possible model prediction without assuming 
a specific form for the confining forces. IK-h.o. are the IK model predictions, where a harmonic oscillator basis is assumed.  }
\label{table1} 
\end{table}

Finally, there are also relations among hadronic parameters which do not
involve the $\Delta$ states. These relations depend also on the
splittings within the $J=1/2,3/2$ pair of states, defined as 
\begin{eqnarray}
\Delta N_{1/2} &=& N(1535) - N(1650) \,, \\ 
\Delta N_{3/2} &=& N(1520) - N(1700) \,.
\end{eqnarray}

There are three such relations: 
\begin{eqnarray} (I) &:& -\frac{5}{18}
\Delta N_{1/2} \cos 2\theta_{N1} 
- \frac29 \Delta N_{3/2} \cos 2\theta_{N3} = N_{5/2} - \frac59 \langle
  N_{1/2}\rangle
- \frac49 \langle N_{3/2}\rangle \,, \\ 
(II) &:& \frac12 \Delta N_{1/2} \cos
  2\theta_{N1} 
- \frac12 \Delta N_{3/2} \cos 2\theta_{N3} = - \langle N_{1/2}\rangle +
  \langle N_{3/2}\rangle \,, \\ 
\label{III} (III) &:&  \Delta N_{1/2} \sin
2\theta_{N1} + \sqrt{10} \Delta N_{3/2} \sin 2\theta_{N3}=0 \,.
\end{eqnarray} 

Any two of these equations fix the mixing angles
$(\theta_{N1},\theta_{N3})$, with different results for the three ways
of choosing two equations. In particular, the first two equations give
\begin{eqnarray}\label{I} 
\Delta N_{1/2} \cos 2\theta_{N1} &=& \frac29
\langle N_{1/2}\rangle + \frac{16}{9} \langle N_{3/2}\rangle  - 2N_{5/2} \,,
\\ 
\label{II} \Delta N_{3/2} \cos 2\theta_{N3} &=& \frac{20}{9} \langle
N_{1/2}\rangle 
- \frac{2}{9} \langle N_{3/2}\rangle  - 2N_{5/2} \,.  
\end{eqnarray}
 
\begin{figure}[t!]
\includegraphics[width=8.0cm]{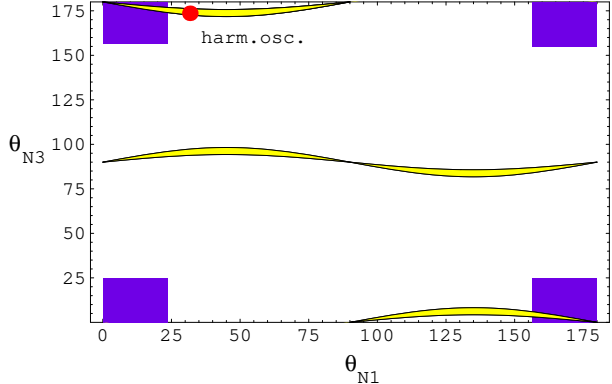} \caption{Constraint on the mixing
angles $(\theta_{N1},\theta_{N3})$ in the general IK model, without any
assumptions about the spatial wave functions.  The four rectangles give
the constraints from Eqs.~(\ref{I}), (\ref{II}), and the yellow bands
represent the constraint Eq.~(\ref{III}). The red dot shows the mixing
angles Eq.~(\ref{mixharm}) obtained in the IK model with harmonic oscillator 
wave functions.} \label{fig:IK} \end{figure}

These equations give $\cos 2\theta_{N1} = 1.081 \pm 0.401,  
\cos 2\theta_{N3}= 0.889 \pm 0.246$, which leads to the allowed
 ranges for the mixing
angles $\theta_{N1} = (0^\circ, 23.6^\circ), (156.4^\circ, 180^\circ)$,
and $\theta_{N3} = (0^\circ, 25.0^\circ), (155.0^\circ, 180^\circ)$.
These ranges are shown in Fig.~\ref{fig:IK} as  rectangles, along
with the constraint from Eq.~(\ref{III}) (the yellow bands). The three
constraints intersect in the upper left and lower right corners of the figure.

The results for the mixing angles in the upper left region are close to the values determined
from $N^*\to N\pi$ strong decays \cite{Hey:1974nc}. The analysis of the
strong decays in Ref.~\cite{Goity:2004ss} gave $(\theta_{N1}, \theta_{N3})=
(22.3^\circ, 136.4^\circ)$ and $(22.3^\circ, 161.6^\circ)$. The second
point is favored by a $1/N_c$ analysis of the photoproduction amplitudes
in Ref.~\cite{Scoccola:2007sn}.

In a recent paper \cite{Pirjol:2008gd} we presented the determination of 
the mixing angles in the one-gluon exchange (OGE) model, where we allow for 
a more general spatial dependence of the hyperfine interaction and also 
include the spin-orbit interaction.  We comment on these
results briefly, since the Isgur-Karl model considered here is a 
limiting case of the OGE model. Considering only the nonstrange states,
the mixing angles of the OGE model are in agreement, within errors, with 
those extracted from strong decays; however, the predicted SU(3) splitting 
$\Lambda_{3/2}(1520)-\Lambda_{1/2}(1405)$ is in disagreement with the observed splitting. 
To correctly reproduce the splitting of these states one also needs flavor dependent operators 
\cite{Schat:2001xr} that partially cancel out the the spin-orbit interaction coming from the one-gluon exchange 
interaction. 

Finally, we quote briefly the best fit values  for the coefficients $c_0,a,b,c$ 
\begin{eqnarray} 
c_0 &=& 1368\pm 11 \mbox{ MeV}\,, \nonumber \\ 
a &=& 155 \pm 8 \mbox{ MeV} \,, \nonumber \\ 
b &=& - 4^{+9}_{-10} \mbox{ MeV} \,, \nonumber \\
c &=& - 8^{+11}_{-12} \mbox{ MeV} \,.
\end{eqnarray}
The resulting masses are listed in Table \ref{table1} as IK-V(r). The fit to the seven masses with four coefficients has three degrees of freedom. The resulting chi squared by degree of freedom 
is $\chi^2_{dof}=1.7$.

\section{The Isgur-Karl model with harmonic oscillator wavefunctions}
\label{IKho}

In the usual treatment of the IK
model \cite{Isgur:1977ef}(denoted here as IK-h.o.), 
the leading order Hamiltonian $H_0$ describes three
constituent quarks interacting by harmonic oscillator potentials
\begin{eqnarray} 
H_0 = \frac{1}{2m} \sum_i p_i^2 + \frac{K}{2}
\sum_{i<j} r_{ij}^2 \,, 
\end{eqnarray} 
This can be diagonalized exactly in
terms of the reduced coordinates $\vec \rho = \frac{1}{\sqrt2}(\vec r_1
- \vec r_2), \vec \lambda = \frac{1}{\sqrt6}(\vec r_1 + \vec r_2 - 2
\vec r_3)$.

Expressed in terms of these coordinates, the Hamiltonian takes the form 
of two independent oscillators
\begin{eqnarray} 
H = \frac{p_\rho^2}{2 m} + \frac{p_\lambda^2}{2 m} +
\frac32 K \rho^2 + \frac32 K \lambda^2 \,.
\end{eqnarray} 
The eigenstates are $\Psi^{\rho,\lambda}_{Lm}$ with
$L=1,m=1$ are 
\begin{eqnarray} \label{psilambda1}
\Psi^\rho_{11} &=& \rho_+
\frac{\alpha^4}{\pi^{3/2}}\exp\left(-\frac12 \alpha^2 (\rho^2 +
\lambda^2)\right) \ , \\ 
\Psi^\lambda_{11} &=& \lambda_+
\frac{\alpha^4}{\pi^{3/2}}\exp\left(-\frac12 \alpha^2 (\rho^2 +
\lambda^2)\right) \,, \label{psilambda2}
\end{eqnarray} 
where $\alpha = (3 K m)^{1/4}$, $\rho_+ = \rho_x + i \rho_y$, $\lambda_+ = \lambda_x + i \lambda _y$ and the combination 
$\rho^2 + \lambda^2$ is invariant under permutations of the three quarks.

The relation to the $\chi$ basis in Eq.~(\ref{chibasis}) is 
\begin{eqnarray}\label{chi2lambdapsi}
\chi_2(1m) &=& \sqrt2 \Psi^\rho_{1m} , \ \\
\chi_3(1m) &=& \frac{1}{\sqrt2} \Psi^\rho_{1m} + \sqrt{\frac32} \Psi^\lambda_{1m} \ .  
\end{eqnarray}
It is easy to check that these states transform under permutations
as specified by the relations Eqs.~(\ref{chibasis}), and are also normalized
correctly.  

The reduced matrix elements of the orbital operators $\langle R_S\rangle, 
\langle Q_S\rangle, \langle Q_{MS}\rangle$ can be computed explicitly using the
wave functions Eqs.~(\ref{psilambda1}),(\ref{psilambda2}), where the expression for the 12 component of a  general spatial operator, 
Eq.~(\ref{calR12}), takes the diagonal form
\begin{eqnarray}
\label{Rdiag}
\langle \Psi_i |{\cal R}_{12} |\Psi_j \rangle &=&
\frac13 
\left( 
\begin{array}{cc} 
\langle {\cal R}_{S} \rangle+\langle {\cal R}_{MS} \rangle &  0  \\ 
  0                                                        & \langle {\cal R}_{S} \rangle-\langle{\cal R}_{MS} \rangle \\ 
\end{array} 
\right)_{ij} \,.
\end{eqnarray}
It is easy to understand that the off-diagonal matrix elements of ${\cal R}_{12}$ (which is symmetric under $P_{12}$)
are zero because $\rho$ and $\lambda$ are antisymmetric and symmetric under $P_{12}$ respectively.

The reduced matrix element $\langle R_S\rangle$ of the spin-spin interaction
can be extracted by considering the matrix element
\begin{eqnarray}
& &\langle \Psi^\lambda_{11} | \delta^{(3)}(\vec r_{12}) |\Psi^\lambda_{11} \rangle = 
A\frac{\alpha^8}{(2\pi)^{3/2}}
\int d^3 \rho \ d^3 \lambda \
\delta^{(3)}(\vec \rho) (\lambda_1^2 + \lambda_2^2) 
e^{-\alpha^2(\rho^2+\lambda^2)} = 
\frac{\alpha^3}{(2\pi)^{3/2}}
\end{eqnarray}
which using the definition of $R_{12}$, Eq.~(\ref{defRQ}), gives
\begin{eqnarray}
\langle R_S\rangle = A\frac{2\alpha^3}{\sqrt{2\pi}} \equiv \delta \,.
\end{eqnarray}
It is convenient to define the parameter $\delta$ as all the other reduced matrix elements 
can be written in terms of this single parameter. 

The computation of the reduced matrix elements for the tensor interaction 
$\langle Q_S\rangle, \langle Q_{MS}\rangle$ is more involved. The analog of Eq.~(\ref{R12}) for the matrix element of the tensor interaction 
$Q_{12}^{ab}$ acting on the quarks 1,2 is given by Eq.~(\ref{Q12}).

The reduced matrix elements $\langle Q_S\rangle$ and $\langle Q_{MS}\rangle$
can be determined from the matrix elements of $Q_{12}^{ab}$ on the 
$\Psi^\lambda, \Psi^\rho$ states. In this basis the matrix element of $Q_{12}^{ab}$
is diagonal as in Eq.~(\ref{Rdiag}). The dependence on the angular momentum projections 
(shown in Eq.~(\ref{Q12}) ) is easy to compute by choosing $a=b=3$,
which gives 
$(\frac12 \{ L^3, L^3\} - \frac23)_{m'=1,m=1} = \frac13$.
The two matrix elements we need are
\begin{eqnarray}\label{melambda}
& &\langle \Psi_{11}^\lambda | Q_{12}^{33} |\Psi_{11}^\lambda \rangle = 0 \,, \\
\label{merho}
& &\langle \Psi_{11}^\rho | Q_{12}^{33} |\Psi_{11}^\rho \rangle 
= - A\frac{4\alpha^3}{15\sqrt{2\pi}} = -\frac{2}{15} \delta \, .
\end{eqnarray}

The first relation can be understood intuitively as following from the
fact that the orbital angular momentum of the quarks 1,2 in the $\Psi^\lambda$
state vanishes, $L_\rho=0$. The tensor operator $Q_{12}^{ab}$ has $L_\rho=2$ and thus its 
matrix element on these states vanishes. Explicitly, the matrix element is
expressed as an integral over $\vec \rho, \vec \lambda$ as
\begin{eqnarray}
\langle \Psi_{11}^\lambda | Q_{12}^{33} |\Psi_{11}^\lambda \rangle = 
A\frac{\alpha^8}{2^{3/2} \pi^3}
\int d^3\rho \ d^3\lambda \
\frac{1}{\rho^5} (3\rho_3^2-\rho^2) (\lambda_1^2 + \lambda_2^2)
e^{-\alpha^2(\rho^2+\lambda^2)} = 0\,,
\end{eqnarray}
since the angular $\rho$ integration vanishes
$\int_{-1}^1 d\cos\theta (3\cos^2\theta-1) = 0$.

The matrix element in Eq.~(\ref{merho}) can be computed
straightforwardly with the result
\begin{eqnarray}
\langle \Psi_{11}^\rho | Q_{12}^{33} |\Psi_{11}^\rho \rangle = 
 A\frac{\alpha^8}{2^{3/2} \pi^3}
\int d^3\rho \ d^3\lambda \
\frac{1}{\rho^5} (3\rho_3^2-\rho^2) (\rho_1^2 + \rho_2^2)
e^{-\alpha^2(\rho^2+\lambda^2)} = - A \frac{4\alpha^3}{15\sqrt{2\pi}}\,.
\end{eqnarray}

Comparing the results with  Eq.~(\ref{Rdiag}),  one finds that the
reduced matrix elements  in the IK model with harmonic oscillator
wave functions are all related and can be expressed in terms of the single parameter $\delta$ as
\begin{eqnarray} 
\langle Q_{MS}\rangle = \langle Q_S\rangle = -
\frac35 \delta \qquad ; \qquad \langle R_S\rangle = \delta \,.
\end{eqnarray} 
This gives a relation among
the coefficients $a,b,c$ of the mass matrix Eq.~(\ref{IKMass}) 
\begin{eqnarray} \label{coefIK}
a = \frac12 \delta \,, \qquad b = \frac{1}{20} \delta \,, \qquad 
c = - \frac15 \delta \,.
\end{eqnarray}

We recover the well known result that in the harmonic oscillator model,  the entire spectroscopy of 
the $L=1$ baryons is fixed by one
single constant $\delta$, along with an overall additive constant $c_0$, and
the model becomes very predictive. The explicit mass matrix is
\begin{eqnarray}\label{IKfirst} 
M_{1/2} &=& (c_0 + \frac34 \delta) + \frac14 \delta \left(
\begin{array}{cc} 
-1 &  -1 \\ 
-1 &   0\\ 
\end{array} 
\right) \,, \\ 
M_{3/2} &=& (c_0 + \frac34 \delta) + \frac14 \delta \left( 
\begin{array}{cc} 
-1 & \frac{1}{\sqrt{10}} \\ 
\frac{1}{\sqrt{10}} & \frac95 \\ 
\end{array}
\right) \,, \\ 
M_{5/2} &=& (c_0 + \frac34 \delta) + \frac15 \delta \,,  \\ 
\Delta_{1/2}
&=& \Delta_{3/2} = (c_0 + \frac34 \delta) + \frac14 \delta \,. 
\label{IKlast} 
\end{eqnarray} 
This
agrees with the mass matrix  of Ref.~\cite{Isgur:1977ef}. Furthermore, the agreement on the signs of the
mixing terms indicates that the phase convention of the states in
Ref.~\cite{Isgur:1977ef} is the same as the phase convention of
Ref.~\cite{Carlson:1998vx} used here.

The mixing angles are independent of the hadron masses, and are given by
\begin{eqnarray}\label{mixharm} 
\theta_{N1} = \arctan (\frac12 (\sqrt5 -
1)) = 31.7^\circ\,,\qquad \theta_{N3} = \arctan
(-\frac{\sqrt{10}}{14+\sqrt{206}}) = 173.6^\circ\,.  
\end{eqnarray} 
The arguments of the previous section show that this prediction is specific
to the harmonic oscillator model. However, the more general predictions of the
IK-V(r) model for the mixing angles are close to this result, as can be seen
from Fig.~\ref{fig:IK}, where the point given in Eq.~(\ref{mixharm}) is indicated as
the red dot.

\section{Relation to the $1/N_c$ expansion}
\label{ncexp}

The predictions of the nonrelativistic quark model can be understood from 
QCD within the large $N_c$ expansion. This method relies on a
power counting scheme to organize the contributions of the different operators according to their order in $1/N_c$.  At leading order in $1/N_c$ the spin-flavor
contracted symmetry $SU(4)_c$ emerges in the baryon sector of QCD \cite{Dashen:1993jt}. 
In the ground state baryon sector, the predictions of this symmetry
reproduce the spin-flavor relations of the constituent quark model. 

The
situation is more complicated for the excited baryons, where the leading
$N_c$ predictions of the contracted symmetry do not generally agree with
those of the quark model \cite{Goity:1996hk,Pirjol:1997bp,Pirjol:2003ye}. 
For example, at leading order in $1/N_c$ the
masses of the non-strange $L=1$ negative parity baryons form three groups of degenerate states
(towers), which differs from the quark model prediction of a degenerate
${\bf 20}$ multiplet of $SU(4)$ \cite{Pirjol:1997bp,Pirjol:2003ye}.

The mass operator of the IK model, Eq.~(\ref{IKMass}),  matches
a subset of the operators that appear in the systematic $1/N_c$ expansion.
 The complete basis was given in Ref.~\cite{Carlson:1998vx} and it
includes core and excited quark operators. The operators
$S_c^2$ and $L_2^{ia} \{s_1^i, S_c^j\}$ contribute at order $O(1/N_c)$,
and the operator $L_2^{ia} \{S_c^i, S_c^j\}$ appears only at  order $O(1/N_c^2)$. 
Using the notation of Ref.~\cite{Carlson:1998vx} the predictions of the IK 
model encoded in Eq.~(\ref{IKMass}) (supplemented by the relations Eq.~(\ref{coefIK}) in the particular case of the IK-h.o. model), can 
be rewritten as
\begin{eqnarray} \label{mIKNc}
H^{eff} &=& c_1 O_1 + c_6 O_6 + c_8 O_8 + c_{17} O_{17}  \\
&=& c_1 N_c 1  + c_6 \left(\frac{1}{N_c} S_c^2\right)  
+ c_8 \left(\frac{1}{2 N_c} L_2^{ab} \{s_1^a\,, S_c^b\}\right)  
+ c_{17} \left(\frac{1}{2 N_c^2} L_2^{ab} \{ S_c^a\,, S_c^b\}\right) \,. \nonumber
\end{eqnarray}
These coefficients are related to the coefficients $c_0,a,b,c$ used in Section \ref{IKV} as 
\begin{eqnarray} \label{cIK1}
c_1 &=&  \frac13 c_0 =  \frac13 m_0 - \frac14 \delta = 462 \ {\rm MeV} \,, \\ 
c_6 &=&  3 a = \frac32 \delta = 450 \ {\rm MeV} \,, \\ 
c_8 &=&  6 c =  -\frac65 \delta = -360\ {\rm MeV} \,, \\ 
c_{17} &=& 18 b = \frac{9}{10} \delta = 270\  {\rm MeV} \,,\label{cIK17}
\end{eqnarray}
where $m_0 = 1610 \ {\rm MeV}$ and $\delta=300 \ {\rm MeV}$ in the IK-h.o. model.
In Table~\ref{table2} the coefficients are compared with the result of 
the best fit made in Section~\ref{IKVpred}. The success of the IK-h.o. 
basically lies in the correct prediction of the value of $c_6$ and the dominance 
of the operator $O_6$ in the general expansion. The predicted
values for $c_8$ and $c_{17}$ in the IK-h.o. model are too large and spoil the fit.
In the best possible fit these two coefficients are compatible with zero within errors.  

\begin{table}
\begin{tabular}{|ccccc|} 
\hline  
    & $c_1$ & $c_6$ & $c_8$ & $c_{17}$  \\ 
\hline
IK-V(r)& 
\ $456 \pm 3.7$ \ & \ $465\pm 23 $ \ & \ $ -46^{+63}_{-74} $ \ & \ $ -69^{+165}_{-186} $ \  \\ 
\hline
IK-h.o.  &  
$ 462 $ & $450$ & $-360$ & $270$ \\ 
\hline
\end{tabular} 
\caption{The coefficients of the best fit in the IK-V(r) and the predicted values for the coefficients in the IK-h.o. model. }
\label{table2} 
\end{table}

In the IK model  with harmonic oscillator wave functions
 $ \delta$ is also related
to the splitting of the ground state baryons as 
$m_N = m_0' - \delta/2,m_\Delta = m_0' + \delta/2$. A simple calculation shows that the effective hamiltonian 
for the ground state baryons that reproduces these IK predictions is
\begin{eqnarray}
H_{gs}^{eff} = g_1  N_c \mathbf{1}  + g_3 \frac{1}{N_c} S^i S^i \,,
\end{eqnarray}
where 
\begin{eqnarray}
g_1 &=& \frac13 m_0' - \frac14 \delta = \frac{5 M_N - M_\Delta}{12} \sim 287 \ {\rm MeV} \,, \\ 
g_3 &=& \delta = M_\Delta - M_N \sim 300 \ {\rm MeV} \,.
\end{eqnarray}

This explicit example is useful to discuss the alternative approach to the $1/N_c$ expansion for
excited baryons presented in Ref.~\cite{Matagne:2006dj}. The authors of 
 Ref.~\cite{Matagne:2006dj} propose an operator basis that differs
from the one in Ref.~\cite{Carlson:1998vx} in that only a subset of the operators are
allowed. More precisely, only operators which do not depend on the
excited and core quarks are present, namely 
\begin{eqnarray}
Q_1 &=& N_c \mathbf{1}  \,, \\
Q_2 &=& L^i s^i \,, \\
Q_3 &=& \frac{1}{N_c} S^i S^i \,, \\
Q_4 &=& \frac{1}{N_c} T^a T^a \,, \\
Q_5 &=& \frac{15}{N_c} L^{(2)ij} G^{ia} G^{ja} \,, \\
Q_6 &=& \frac{3}{N_c} L^i T^a G^{ia} \,, \\
Q_7 &=& \frac{3}{N_c^2} S^i T^a G^{ia} \,.
\end{eqnarray}
The first observation is that these seven operators are not independent. We find that $Q_7$ can be rewritten 
in terms of $Q_1$, $Q_3$, $Q_4$ as:  $Q_7 = - \frac{3(4 N_c-9)}{16 N_c^3} Q_1 + \frac{3 (N_c-1)}{8 N_c} (Q_3 + Q_4) $.
Furthermore, using the matrix elements from Table 3 in the first of  Refs.~\cite{Matagne:2006dj} and equating $\sum_{i=1}^6 c_i Q_i $ to 
the matrix elements of the Isgur-Karl model, Eq.~(\ref{IKfirst})-(\ref{IKlast}) it is easy to see that it is not 
possible to find coefficients $c_i$ that reproduce the predictions of the IK 
model. This
is an explicit example that shows that the basis proposed in \cite{Matagne:2006dj} is incomplete. 

For the completely symmetric ground state baryons the $\{Q_i\}$ basis is correct, 
but overcomplete, as only $Q_1,Q_3$ are needed. The $\{Q_i\}$ basis constructed with symmetric 
operators is only correct for symmetric spin-flavor states like the $[\mathbf{56},L=2]$, see for example 
Ref.~\cite{Goity:2003ab}.

\section{Conclusions}
\label{concl}

We showed in this paper how to construct the effective mass operator of 
the Isgur-Karl model for the non-strange negative parity $L=1$ excited 
baryons. The effective mass operator is written as an operator expansion 
in Eq.~(\ref{IKMass}), where the spatial dependence and spin-flavor dependence 
are factorized. This form of the mass operator is valid without making any
assumptions about the spatial dependence of the quark wave functions and 
allows to explore the IK model beyond the harmonic 
oscillator approximation. The unknown spatial dependence is contained in 
the three coefficients Eqs.~(\ref{coefa})-(\ref{coefc}) of the expansion 
which are written in terms of orbital overlap integrals in 
Eqs.~(\ref{R12})-(\ref{Q12}). These explicit expressions for the coefficients 
are obtained exploiting the tranformation properties of states and interactions 
under the permutation group $S_3$ acting on the spatial and spin-flavor degrees 
of freedom~\cite{Pirjol:2007ed}. The spin-flavor structure of the model is manifest 
in the three non-trivial  
operators that appear 
in the expansion, whose matrix elements are calculable and can be conveniently 
read off from Tables II and III in Ref.~\cite{Carlson:1998vx}.

The general operator form Eq.~(\ref{IKMass}) 
leads to parameter free mass relations that also constrain the mixing 
angles and are well satisfied by data. The most noticeable disagreement 
is the prediction of the degeneracy of the two $\Delta$ states. The 
experimental data seems to point to the presence of a spin-orbit interaction. 
Smaller experimental errors on the masses of these two states would contribute 
to determine its strength.

In the particular case of harmonic oscillator  wave functions the coefficients 
of the mass operator can be computed and written in terms of a single parameter 
as shown in Eq.~(\ref{coefIK}). The mass operator Eq.~(\ref{IKMass}) reproduces 
then exactly the predictions of the  IK model as formulated in Ref.~\cite{Isgur:1977ef}. 
As is well known, in this approximation the mixing angles are fixed, independently 
of the hadronic parameters.

Recasting the predictions of the  IK model in this way makes clear its relation to
the $1/N_c$ studies of excited baryons, where the spin-flavor 
quark operator expansion is used in a systematic way. In Eq.~(\ref{mIKNc})
and Eqs.~(\ref{cIK1})-(\ref{cIK17}) we present the result of the matching of the IK 
model to the operators of the $1/N_c$ expansion, using the notation of 
Ref.~\cite{Carlson:1998vx}. 

The matching of the IK model is a simple example that shows that 
operators depending on the excited quark and core quark decomposition are 
necessary~\cite{Pirjol:2007ed}. The alternative operator basis proposed 
in Ref.~\cite{Matagne:2006dj} which does not include core operators can 
not reproduce the mass operator of the IK model with harmonic oscillator 
wave functions, and is thus incomplete.

\begin{acknowledgments}
The work of C.S. was supported by CONICET and partially supported by  the U.~S.  Department of Energy, Office of
Nuclear Physics under contract No. DE-FG02-93ER40756
with Ohio University. 
\end{acknowledgments}

\end{document}